\documentclass[aps,prl,twocolumn,groupedaddress,showpacs]{revtex4}
\usepackage{graphicx}
\usepackage{dcolumn}
\usepackage{bm}
\usepackage{amsmath,amsfonts,amssymb,amsthm,verbatim}
\usepackage{floatflt}
\usepackage[ansinew]{inputenc}
\usepackage[usenames]{pstcol}

\usepackage{color}
\usepackage{epsfig}

\newcommand{\be}{\begin{equation}}
\newcommand{\ee}{\end{equation}}
\newcommand{\bea}{\begin{eqnarray}}
\newcommand{\eea}{\end{eqnarray}}
\newcommand{\beq}{\begin{eqnarray}}
\newcommand{\eeq}{\end{eqnarray}}

\newcommand{\bi}{\begin{itemize}}
\newcommand{\ei}{\end{itemize}}
\newcommand{\bc}{\begin{center}}
\newcommand{\ec}{\end{center}}

\newcommand{\up}{\uparrow}
\newcommand{\down}{\downarrow}

\newcommand{\la}{\langle}
\newcommand{\ra}{\rangle}

\newcommand{\reff}{\text{eff}}

\begin{document}

\title{Metal-insulator transition in a quantum wire driven by a modulated Rashba spin-orbit coupling}
\author{G. I. Japaridze$^1$, Henrik Johannesson$^{2,3}$, and Alvaro Ferraz$^4$}
\affiliation{$\mbox{}^1$ Andronikashvili Institute of Physics, Tamarashvili 6, 0177 Tbilisi, Georgia}
\affiliation{$\mbox{}^2$ Department of Physics, University of
Gothenburg, SE 412 96 Gothenburg, Sweden}
\affiliation{$\mbox{}^3$ Kavli Institute for Theoretical Physics, University of California,
Santa Barbara, California 93106-4030}
 \affiliation{$\mbox{}^4$ ICCMP, University of Brasilia, 70904-910 Brasilia-DF, Brazil}

\begin{abstract}  
We study the ground-state properties of electrons confined to a quantum wire and subject to a smoothly modulated Rashba spin-orbit coupling. When the period of the modulation becomes commensurate with the band filling, the Rashba coupling drives a quantum phase transition to a nonmagnetic insulating state. Using bosonization and a renormalization group approach, we find that this state is robust against electron-electron interactions. The gaps to charge- and spin excitations scale with the amplitude of the Rashba modulation with a common interaction-dependent exponent.  An estimate of the expected size of the charge gap, using data for a gated InAs heterostructure, suggests that the effect can be put to practical use in a future spin transistor design.

\end{abstract}

\maketitle

{\bf Introduction $-$} Progress in the control and
manipulation of spin degrees of freedom in semiconductors holds
great promise for the development of future spintronics devices
\cite{Review}. Much of current work in the field is inspired by
various proposals for spin transistors. In what has become the
prototype for a spintronics device scheme $-$ the Datta-Das spin
transistor \cite{DattaDas} $-$  spin-polarized
electrons are injected from a ferromagnetic source into a quasi
one-dimensional (1D) ballistic channel (''quantum wire'') formed in
a semiconductor heterostructure. The structure inversion asymmetry
of the heterostructure produces a Rashba spin-orbit coupling that
makes the spins of the electrons precess with a rate controllable 
via a gate. In a simplified picture, an electron with the
same spin projection as that of the magnetized drain is accepted by
the drain, or else the electron is scattered away. This realizes an
''on-off'' current switch, controllable by the gate bias. The scheme is
yet to be realized, however. One obstacle is the inefficiency of
present techniques for injecting spin-polarized electrons from a
ferromagnet into a quantum wire. Alternative ideas for developing 
a current switch based on a Rashba coupling 
are thus in high demand. 

In an analysis of 1D spin transport, Wang \cite{Wang} suggested a design for a
spin transistor where the electrons experience a spatially periodic
Rashba spin-orbit coupling.  In this scheme segments of
a quantum wire with a uniform Rashba coupling are connected in
series to segments with no coupling. A Fabry-P\'{e}rot-like interference 
between electron waves scattered at the interfaces between two segments 
leads to a transmission gap with a complete
blocking of the charge current over a range of energies when the
number of segments becomes sufficiently large. By tuning the electron
density $-$ and hence the Fermi level $-$ by a supplementary gate,
the flow of current in the wire can then be controlled {\black effectively}. 
As pointed out by Gong and Yang \cite{GongYang}, the
effect is fully operative for electrons with no spin polarization.
By utilizing a periodically modulated Rashba coupling one may thus
envision a spin transistor without the injection of spin-polarized
electrons into the current-carrying channel \cite{Kirczenow}.

This intriguing prospect motivates a closer investigation. In the
present paper we address two issues: first, how robust is the
opening of a charge excitation gap against smoothening of the
boundaries between regions with different strengths of the Rashba
coupling? In particular, if the Rashba strength varies
continuously on the scale of the underlying lattice, can a gap still
appear? If so, under what conditions? Secondly, how
do electron-electron interactions influence the gap-opening? This
question is crucial in view of applications, as electron
interactions in 1D can dramatically change the physics expected from
an independent-electron picture \cite{Giamarchi_book_04}. As we shall
see, by "locking" the band filling to the periodicity of the Rashba modulation,
a gap to charge excitations $-$ as well as to spin excitations $-$ does open up for 
a smooth Rashba interaction {\black and it persists even when} electron interactions are included. This gap-opening mechanism is very different from that based on repeated potential scattering in Refs. \onlinecite{Wang} and \onlinecite{GongYang}, the only common ingredient being the presence of a periodic Rashba modulation.
In fact, we find that {\black in the experimentally relevant parameter range}, {\em electron interactions  increase the size of the charge gap}, 
thus assisting the use of a gate-controlled modulated Rashba coupling as a current switch. 

{\bf Non-interacting electrons $-$}  We consider a Rashba spin-orbit
interaction $H_R$, which can be split into a uniform and a harmonically varying piece,
\begin{equation} \label{Rashba}
{\black H_R = \left(\alpha_0 k_x  + \frac{\alpha_1}{2} \{\cos(Q x), k_x\}\right) \sigma^y.}
\end{equation}
Here $\alpha_0$ and $\alpha_1$ are constants, $Q$ is a wave number,
$k_x$ is the electron wave number along the wire, and $\sigma^y$ is
a Pauli matrix. {\black The anticommutator $\{\cos(Q x), k_x\}$ ensures that the interaction is
Hermitian.} The structure of Eq. (\ref{Rashba}) may be used in an
attempt to qualitatively capture the effect of a {\black piecewise} modulated 
Rashba coupling in a quantum wire where distortions and stray electric
fields smoothen the sharp interface between two consecutive segments
of the wire (each of extension $l_{0} = 2\pi/Q$) 
with different values of the coupling. 
The real {\em raison d'\^etre} for
our choice in Eq. (\ref{Rashba}), however, is that it allows for a
well-controlled analysis of a modulated Rashba coupling,
also in the presence of electron interactions.

{\black To set the stage, let us first focus on the case of non-interacting electrons}. We shall assume 
that only the lowest-energy sub-band is partly filled, as this is the case most
relevant for an experimental realization.
Making use of a {\black tight-binding} lattice formulation, we represent the kinetic energy
by 
\begin{equation} \label{TightBinding}
H_0 = -t \sum_{n, \mu}
\left(c_{n,\mu}^{\dag}c^{\phantom{\dagger}}_{n+1,\mu} + \mbox{H.c.} \right).
\end{equation}
Here $t$ is the hopping amplitude, and $c^{\dagger}_{n, \mu}$ $(c^{\phantom{\dagger}}_{n,\mu})$
are electron creation (annihilation) operators on site $n$ with
spin projection $\mu = \uparrow, \downarrow$ along the $\hat{z}$-axis. The role of the Rashba interaction in Eq. (\ref{Rashba}) is taken by
\begin{equation}  \label{LatticeRashba}
H_R \! = \! -i \sum_{n, \mu, \nu} \! (\gamma_0+\gamma_1 \cos \left(Q na)\right) \! \left(\!
c_{n, \mu}^{\dag}\sigma^y_{\mu \nu} c^{\phantom{\dagger}}_{n+1, \nu} \! - \!
\mbox{H.c.}\,\! \right)\!
\end{equation}
where $\gamma_j = \alpha_j a^{-1} \ (j\!=\!0,1)$, with $a$ the
lattice spacing. It is useful to introduce spin-rotated
operators  $b^{\phantom{\dagger}}_{n, + }\! \equiv \! (c_{n,\down} -
i \, c_{n,\up})/\sqrt{2}$ and $b^{\phantom{\dagger}}_{n,-} \!
\equiv\! (c_{n,\up} -  i \, c_{n,\down})/\sqrt{2}$, and {\black write} the
Hamiltonian $H = H_0 + H_R$ as
\begin{multline}  \label{RotatedHamiltonian}
H  =  - \sum_{n, \tau}
\!\Big[\! \,(t+i\tau\gamma_{0})b^{\dag}_{n,\tau}b^{\phantom{\dagger}}_{n+1,\tau}\!+\!\mbox{H.c.}
\!\,  \\
- i \gamma_{1}\! \cos(Q na) ( \,\tau\,
b^{\dag}_{n,\tau}b^{\phantom{\dagger}}_{n+1,\tau}\!\!-\!\mbox{H.c.})\Big],
\end{multline}
with $\tau = \pm$ labeling the eigenstates of the $\sigma_y$ operator, i.e. the spin projections 
on the axis along which the effective momentum-dependent Rashba field is pointing.
When $\gamma_1=0$, the Hamiltonian
in Eq. (\ref{RotatedHamiltonian}) describes a 1D system of
non-interacting electrons in the presence of a uniform Rashba
spin-orbit coupling. For this case the Hamiltonian is readily
diagonalized in momentum space, and one finds that the
spin-degenerate {\black band}  in the absence of Rashba coupling gets
shifted {\black horizontally} into two {\black distinct} branches,
\begin{equation} \label{Dispersion}
E^{0}_{\tau}(k)  = -2 {\tilde{t}}
\cos\left((k-\tau q_0)a \right),
 \end{equation}
where $q_{0}  =  a^{-1}\arctan \left(\gamma_{0}/t\right)$ and ${\tilde{t}}=\sqrt{t^{2}+\gamma_0^{2}}$. 
Note that we here consider an {\em ideal} 1D quantum wire, thus avoiding the complication
of energy band deformations produced by a spin-orbit interaction in the presence of
a soft transverse confining potential \cite{Moroz}. 

At band-filling $\nu=N_{e}/N_{0}$, with $N_e \, [N_0]$ being the number of
electrons [lattice sites], the system is characterized by the four
Fermi points $k_{F,R}^{\tau} = k_{F}^{0}\, + \tau\, q_{0},
k_{F,L}^{\tau} =  - k_{F}^{0} + \, \tau q_{0} \ (\tau = \pm)$, where
$k_{F}^{0}= \pi \nu/2a$. To simplify the analysis
we linearize the spectrum around these Fermi points and pass to a continuum limit with 
$na \rightarrow x$. 
By decomposing the lattice operators $b_{n, \tau}$ into right- and
left-moving fields $R^{\phantom{\dagger}}_{\tau}(x)$ and $L^{\phantom{\dagger}}_{\tau}(x)$,
\begin{displaymath}  \label{Decomposition}
b^{\phantom{\dagger}}_{n,\tau}  \rightarrow  \sqrt{a}
\big(\mbox{e}^{i(k_{F}^{0}+\tau q_{0})x} R^{\phantom{\dagger}}_{\tau}(x) +
\mbox{e}^{-i(k_{F}^{0}-\tau q_{0})x}L^{\phantom{\dagger}}_{\tau}(x) \big),
\end{displaymath}
the lattice Hamiltonian in Eq. (\ref{RotatedHamiltonian}) takes the form {\black $H=\int dx\left({\cal H}_{+}+{\cal H}_{-} \right)$}, with
\begin{eqnarray}  \label{linear-Hamiltonian}
&&{\cal H}_{\tau} \!= \!-iv_F
\big(\!:\!R^{\dag}_{\tau}(x)\partial_{x}R^{\phantom{\dagger}}_{\tau}(x)\!:-
:\!L^{\dag}_{\tau}(x)\partial_{x}L^{\phantom{\dagger}}_{\tau}(x)\!:\!\big) \nonumber\\
&&\hspace{-3mm}-2\Delta_R \cos(Q x) \big( \mbox{e}^{-2ik_F^0(x +
a/2)}R^{\dagger}_{\tau}(x)L^{\phantom{\dagger}}_{\tau}(x) \!+
\!\mbox{H.c.}\big),
\end{eqnarray}
where $v_{F}=2a\sqrt{t^{2}+\gamma_{0}^{2}}$ and {\black $\Delta_R =\gamma_1 \sin(q_0a)$}, and where
we have omitted rapidly oscillating terms, which vanish upon integration.
The normal ordering $:...:$ is carried out with respect to the filled Dirac sea.  

{\bf Bosonization.} To make progress we bosonize the theory, using
$R^{\phantom{\dagger}}_{\tau}(x)=\eta_{\tau} \exp \left({\it
i}\sqrt{\pi}[\varphi_{\tau}(x)+\vartheta_{\tau}(x)]\right)/\sqrt{2\pi
a}$ and $L^{\phantom{\dagger}}_{\tau}(x)= \bar{\eta}_{\tau} \exp
\left( -{\it
i}\sqrt{\pi}[\varphi_{\tau}(x)-\vartheta_{\tau}(x)]\right)
/\sqrt{2\pi a}  \label{L}$, where $\varphi_{\tau}(x)$ and
$\vartheta_{\tau}(x)$ are dual bosonic fields satisfying $\partial_t
\varphi_{\tau} = v_{F} \partial_x \vartheta_{\tau}$, and where
$\eta_{\tau}$ and $\bar{\eta}_{\tau}$ are Klein factors which keep
track of the fermion statistics for electrons in different branches
\cite{GNT_book_98}. Inserting the bosonized forms of
$R^{\phantom{\dagger}}_{\tau}(x)$ and
$L^{\phantom{\dagger}}_{\tau}(x)$ into Eq. (\ref{linear-Hamiltonian}),
and introducing the charge $(c)$ and spin $(s)$ fields $\varphi_c
\equiv (\varphi_{+} + \varphi_{-})/\sqrt{2}$ and $\varphi_s \equiv
(\varphi_{+} - \varphi_{-})/\sqrt{2}$, we arrive at the bosonized
Hamiltonian
\begin{multline} \label{Hboson}
H = \int dx \Big(\frac{v_F}{2} \sum_{i=c,s}  \big(
(\partial_{x}\varphi_{i})^2 +
(\partial_{x}\vartheta_{i})^2 \big) \,\\
- \frac{2\Delta_R}{\pi a}\!\sum_{j=\pm 1}\! \sin
([Q+2jk^{0}_{F}]x+k^{0}_{F}a+ \sqrt{2\pi}\varphi_c)
\cos(\sqrt{2\pi}\varphi_s) \Big). \nonumber
\end{multline}
{\black When
$Q-2k^{0}_{F} \simeq {\cal O}(1/a)$, {\black both Rashba terms $\sim
\Delta_{R}$} are rapidly oscillating and average to zero.} 
Thus, in this limit the model {\black describes free charge and 
spin bosons, i.e.} a metallic phase with gapless {\black spin} excitations. 

{\black In contrast, when} $Q-2k^{0}_{F} \ll {\cal O}(1/a)$ the {\black $j=-1$ component of
the} modulated Rashba coupling comes into play. For this case it is useful to perform a 
{\black transformation}, $(Q-2k^{0}_{F})x +k^{0}_{F}a+\sqrt{2\pi}\varphi_{c}
\rightarrow \pi/2 + \sqrt{2\pi}\varphi_{c}$, and rewrite the
Hamiltonian density as
\begin{eqnarray} \label{Bos-Ham-Arbitr-nu}
{\cal H} & = & \frac{v_F}{2} \sum_{i=c,s}  
((\partial_{x}\varphi_{i})^2 +
(\partial_{x}\vartheta_{i})^2 ) 
- \mu_{\reff}\partial_{x}\varphi_{c} \nonumber \\
& - & \frac{2\Delta_R}{\pi a}
\cos(\sqrt{2\pi}\varphi_{c})\cos(\sqrt{2\pi}\varphi_{s}),
\end{eqnarray}
where $\mu_{\reff} \equiv v_{F}\sqrt{2/\pi}(Q-2k^{0}_{F})$ {\black is
an effective  ''chemical potential'', which, when tuned to zero, 
"locks" the band filling to commensurability with the Rashba modulation.}
{\black For this case, i.e. with $\mu_{\reff}=0$,}
the Hamiltonian describes two bosonic charge and spin fields
coupled by the strongly [renormalization-group (RG)] relevant operator
{\black $\cos(\sqrt{2\pi}\varphi_{c})\cos( \sqrt{2\pi} \varphi_{s})$}. This
operator pins the charge and spin fields at their ground
state expectation values
\bea \la \varphi_{c} \ra & = & \la \varphi_{s}\ra
=\sqrt{\pi/2}n\, \qquad n=0,\pm 1,\pm 2,...\,. \label{SGs} \eea
and as a result both spin and charge excitations develop a gap 
\cite{SSH}. Thus, {\em when $\mu_{\reff}=0$, the system
turns into a nonmagnetic insulator.}

{\black To study the properties of the insulating state, specifically the size
of the charge excitation gap, we use a}
mean-field decoupling of charge and spin in Eq.
(\ref{Bos-Ham-Arbitr-nu}) and write the Hamiltonian as $H=\int
\!dx\left({\cal H}_c + {\cal H}_s\right)$, where (for $i=c,s$)
\begin{equation}  \label{MeanFieldDecoupled}
{\cal H}_i \!= \!\frac{v_F}{2} [(\partial_{x}\varphi_{i})^2 \!+
\!(\partial_x \vartheta_{i})^2] - \frac{m_{i}}{\pi
a}\cos(\sqrt{2\pi} \varphi_{i}), 
\end{equation}
with 
\begin{equation}  \label{MassParameters}
m_{c}\!\equiv\! \Delta_R \langle \cos(\sqrt{2\pi}\varphi_{s})\rangle, \ \
\  m_{s}\!\equiv\! \Delta_R \langle \cos(\sqrt{2\pi}\varphi_{c})\rangle.
\end{equation}
Note that the mean-field decoupling 
is here under control since the 
pinning, Eq.~(\ref{SGs}), implies that field fluctuations are strongly suppressed. 
As seen from Eq. (\ref{MeanFieldDecoupled}), the mean-field theory at 
$\mu_{\reff}\!=\!0$ {\black is equivalent} to
two commuting sine-Gordon (SG) models  with $\beta^{2} = 2\pi$, {\black and with
''bare'' masses defined by Eq. (\ref{MassParameters}).} From
the exact solution of the sine-Gordon model it is known that for this case 
the excitation spectrum is gapped and
consists of solitons and antisolitons with mass $M_{c/s}$
and soliton-antisoliton bound states ("breathers") with the lowest
breather mass also equal to $M_{c/s}$ \cite{DHN}. 
As $M_c$
determines the charge gap caused by the modulated Rashba coupling,
we shall derive an expression for $M_c$ that allows us to estimate
its size in a given experimental setting.  Before doing so, however,
let us show how the analysis above
can be {\black extended} so as to take into account the electron-electron interactions.  

{\bf Interacting electrons $-$}  {\black Since Umklapp scattering is absent in a ballistic quantum wire, 
one is left with backscattering
$(\sim g_{1 \pm})$, dispersive scattering $(\sim g_{2 \pm})$, and forward scattering $(\sim g_{4 \pm})$, controlled
by 
\begin{multline} \label{Hint}
{\cal H}_{\text{int}} =  g_{1 -}\!:\!R^{\dag}_{\tau}L^{\phantom{\dagger}}_{\tau}L^{\dag}_{-\tau}R^{\phantom{\dagger}}_{-\tau}\!: 
+  \, \tilde{g}_{2 \tau} \!:\!R^{\dag}_{+}R^{\phantom{\dagger}}_{+}L^{\dag}_{\tau}L^{\phantom{\dagger}}_{\tau}\!:   \\
+  \frac{g_{4 \tau}}{2}( :\!R^{\dag}_{+}R^{\phantom{\dagger}}_{+}R^{\dag}_{\tau}R^{\phantom{\dagger}}_{\tau}\!: + 
\, R \leftrightarrow L) \end{multline}
with $\tau \!= \!\pm$ summed over, $\tilde{g}_{2 \tau} \!\equiv\! g_{2 \tau}\!\,-\!\,\delta_{\tau +}g_{1 \tau}$, and
$(+,-) \leftrightarrow (\parallel, \perp)$ in the standard ''g-ology'' notation \cite{Giamarchi_book_04}. 

{\black The strength of the electron interaction in a semiconductor structure is typically much smaller than the band width. For this weak-coupling case the backscattering 
 $\sim g_{1 -}$ is marginally irrelevant and renormalizes to zero at low energies 
(just as for a 1D electron system in the absence of spin-orbit coupling) \cite{GJPB_05}.  From now on
we therefore consider an effective model where the back scattering has been renormalized away. 
The bosonized mean field theory, including
electron interactions, then takes the form $H=\int
\!dx\,[\,{\cal H}_c + {\cal H}_s]$, where (for $i=c, s$)
\begin{equation}
{\cal H}_i \!= \!\frac{v_i}{2} [(\partial_{x}\varphi_{i})^2 \!+
\!(\partial_x \vartheta_{i})^2]  -
\!\frac{m_{i}}{\pi a}\!\cos(\sqrt{2\pi K_i} \varphi_{i}). 
\label{InteractingMeanFieldBoson}
\end{equation}
For weak interactions, $v_i$ and $K_i$ can be explicitly 
parametrized in terms of the amplitudes in Eq.~(\ref{Hint}) \cite{Giamarchi_book_04}. {\black Note that also the bare masses $m_i$ get renormalized
by the interaction, with $\varphi_i \rightarrow \sqrt{K_i}{\varphi}_i$ in Eq. (\ref{MassParameters}).} It is also important to note that
the breaking of spin-rotational invariance by the Rashba interaction implies that 
the {\black RG} fixed-point value of $K_s$ is not slaved to unity \cite{Giamarchi_book_04},
but can take larger values \cite{footnote1}.

{\bf Charge and spin excitation gaps $-$} \  {\black We can now derive an}
expression for the charge excitation gap $-$ identified as the physical soliton
mass $M_c$ in the charge sector of Eq. (\ref{InteractingMeanFieldBoson}) $-$ with the 
electron interactions included.
The mass $M_c$ {\black and the corresponding mass $M_s$ in the spin sector are related to the (bare)
mass parameters $m_c$ and $m_s$, respectively,} by \cite{Al_B_Zamolodchikov_95}
\begin{equation} \label{PhysicalMass}
M_{i} = {\cal C}_1(K_i) \Lambda (m_{i}/\Lambda)^{2/(4-K_{i})},\qquad i=c,s
\end{equation}
where $\Lambda$ is an energy cutoff, and ${\cal C}_1(K_i)\!\equiv\! [\pi\Gamma(1\!-\!K_i/4)/\Gamma(K_i/4)]^{2/(4\!-\!K_i)} [2\Gamma(\xi_i/2)/\sqrt{\pi}\,
\Gamma(1/2 - \xi_i/2)]$, with $\xi_i = K_i/(4-K_i)$. The ground state expectation
values {\black entering the bare masses $m_{i}$ are in turn related} to the physical masses $M_{i}$ by
\cite{Luk_Zam_97}
\begin{equation} \label{ExpValues}
\langle \cos(\sqrt{2\pi K_i}\varphi_{i})\rangle = 
{\cal C}_2(K_i)(M_i/\Lambda)^{K_i/2}, \ \ \ i=c,s
\end{equation}
with ${\cal C}_2(K_i) \!\equiv \! [(1+\xi_i)\pi \Gamma(1-K_i/4)/16 \sin(\pi \xi_i) \Gamma(K_i/4)] \times [\Gamma(1/2+\xi_i/2)\Gamma(1-\xi_i/2)/4 \sqrt{\pi}]^{(K_i/2)-2}[2\sin(\pi\xi_i/2)]^{K_i/2}$.
Combining Eqs. (\ref{MassParameters}), (\ref{PhysicalMass}) and
(\ref{ExpValues}), some elementary algebra yields for the charge excitation gap.
\begin{equation} \label{MassScaling}
M_c = {\cal C}_c \Lambda(\Delta_R/\Lambda)^{2/(4-K_c-K_s)},
\end{equation}
where ${\cal C}_c^{16-4K_c-4K_s}\equiv{\cal C}_1(K_c)^{(4-K_c)(4-K_s)} {\cal C}_2(K_c)^{2K_s}\times \\
{\cal C}_1(K_s)^{(4-K_s)K_s}{\cal C}_2(K_s)^{2(4-K_s)}$. The spin gap $M_s$ is given by the same expression, but with $c \leftrightarrow s$. 

{\black The opening of the charge gap at a band-filling commensurate
with the period of the Rashba modulation leads to a reduction of the ground state 
energy,  and pins the band filling at this value until the chemical potential 
{\black reaches} the bottom of the upper band.
The competition between the chemical potential $\mu$ and the 
commensurability energy drives a continuous quantum phase transition from 
a gapped (insulating) phase at $\mu< \mu_{c}\!=\!M_{c}$ to a gapless (metallic) phase at $\mu >
\mu_{c}$} \cite{C_IC_transition,JNW_1984}.  Such a
transition belongs to the universality class of a commensurate-incommensurate metal-insulator
transition \cite{Giamarchi_book_04}. The critical conductivity $\sigma$, proportional to the doping
of the upper band, scales as 
$\sigma \sim \left(\mu-\mu_{c}\right)^{1/2}$, while the compressibility $\kappa$ diverges as $\kappa \sim
\left(\mu-\mu_{c}\right)^{-1/2}$ before dropping to zero on the insulating side. In the gapless phase the
ground state expectation value $\langle\cos(\sqrt{2\pi
K_c}\varphi_c)\rangle$ vanishes and, as follows from Eq.
(\ref{MassParameters}), {\black this implies that also the bare mass $m_s$ 
vanishes.} As a consequence, the quantum phase transition in the
charge sector at $ \mu =\mu_c$ is accompanied by a
similar transition in the spin sector, with the system showing
Luttinger liquid behavior with gapless spin and charge excitations
for $\mu > \mu_c$. 

\indent {\bf Implications $-$} Our main result, Eq.
(\ref{MassScaling}), {\black boosts the proposal \cite{Wang,GongYang}} that a controllable and modulated Rashba coupling
may serve as a current switch in a quantum wire. 
It is here important to emphasize that our scheme exploits a nontrivial commensurability property, encoded in the condition $Q = 2k_F$, and is hence different from that in Refs. \onlinecite{Wang} and \onlinecite{GongYang}, which is based on a picture of repeated single-particle scattering.
For an implementation one would need a configuration 
of switchable top gates that produce the modulation, as well as a tunable back gate by which 
the band filling can be adjusted. While a {\black challenging quest} in quantum engineering, 
our {\black analysis testifies to the soundness} of the scheme as it shows that {\em a Rashba-induced charge gap is robust against electron-electron interactions}. In fact, as revealed by Eq. (\ref{MassScaling}), {\black in the experimentally relevant range $K_c+K_s < 2$ \cite{Giamarchi_book_04}}, {\black the gap grows with the strength of the electron interactions.} 

As for the size of the gap, we may take as
illustration a quantum wire patterned in an InAs heterostructure, which, 
due to its strong
spin-orbit coupling and large electron mean-free path, is 
favored in spintronics applications
\cite{Review}. {\black Using data for a heterostructure grown by molecular-beam epitaxy \cite{Grundler},
with Rashba parameter $\hbar \alpha \approx 2\times 10^{-11}$ eVm, carrier density $n_e \approx 1 \times 10^{12}$ cm$^{-2}$, effective mass $m^{\ast} \approx 0.4m_e$, and lattice spacing $a \approx 5$ \AA , we have that $\hbar\Delta_R \simeq 4$ meV, taking $\alpha_1 \approx \alpha$. A rough estimate of the charge and spin stiffness parameters, assuming a well-screened interaction \cite{Hausler} with $\epsilon \approx 15 \epsilon_0$ \cite{Yang}, yields that $K_c \simeq 0.6$ and $K_s \simeq 1.2$. With $\Lambda = \hbar v_F/a \approx 0.5$ eV, it follows from Eq. (\ref{MassScaling})} that $M_c \simeq 1\times 10^2$ meV,
corresponding to a threshold voltage of approximately 100 mV in a spin transistor application. 

Using the estimate {\black of} $M_c$ above, the characteristic length scale 
$\xi \sim \hbar v_F/M_c$ at which the gap starts to open up is given by 
$\xi \simeq 5 $ nm. This number fits
easily within the quantum ballistic regime of an InAs quantum wire,
{\black with an estimated mean-free path of 
 $\gtrapprox 1.5 \, \mu$m \cite{Yang},
and one thus expects the gap to open up fully.}
In this context one {\black should realize}
that an implementation of a gate-controlled Rashba modulation {\black with a well-defined
periodicity} is an experimental
challenge, and must probably await {\black further} progress
in device technology. It is here important to note that a
periodic gate bias will modulate also the electron density, thus favoring a
build-up of charge density wave correlations. While this effect is expected to assist the opening 
of the {\black Rashba} gap, its precise influence {\black needs to be studied} in a more complete theory. 

\indent {\bf Summary $-$} {\black To conclude, we have shown that a smoothly modulated Rashba
spin-orbit coupling in a quantum wire drives a commensurate-incommensurate 
metal-insulator transition at a critical value of the band filling. The
charge excitation gap (as well as the
associated spin gap) is found to scale with the amplitude of the Rashba modulation
$\gamma_1$ as $\gamma_1^{2/(4-K_c-K_s)}$, where $K_c$ and $K_s$ are the charge
and spin stiffness parameters that encode electron interaction effects. 
 In a next step one should try to refine the
analysis of the problem, and go beyond the minimal model employed here.
In particular, effects from local variations in the electron density and
from electron back scattering at very low densities \cite{GJPB_05} {\black are important to explore}.
Also, the question of what happens for more general periodic profiles of the Rashba modulation is an important issue. With a more complete theory,  and with {\black advances} in device technology, a low-bias spin transistor based on a switchable and modulated Rashba coupling may well become a reality. 

\noindent {\em Acknowledgments:} We wish to thank T. Nattermann and
Z. Ristivojevic} for interesting discussions. H.J. acknowledges hospitality of the 
KITP at UCSB. This work was supported 
by STCU (Grant No. 3867)
(G.I.J), the National Science Foundation (Grant No. PHY05-51164)
and Swedish Research Council (Grant No. VR-2005-3942) (H.J.), and 
the Brazilian CNPq and Ministry of Science and Technology (A.F.).

\end{document}